\documentclass[fleqn,twoside]{article}
\usepackage{espcrc2}
\usepackage{amssymb}

\usepackage{graphicx}
\usepackage[figuresright]{rotating}

\newcommand{\AmS}{{\protect\the\textfont2
  A\kern-.1667em\lower.5ex\hbox{M}\kern-.125emS}}

\title{Positivity and Fermionic Dense Matter}

\author{Deog Ki Hong\address
{Department of Physics, Pusan National University,
\\Busan 609-735, Korea}
        and Stephen D. H. Hsu 
\address{Department of Physics, University of Oregon,
\\Eugene OR 97403-5203, U.S.A.}}
\begin{document}

\begin{abstract}
Euclidean dense matter generically suffers from the fermion sign
problem. However, we argue that the sign problem is absent if one
considers only low-energy degrees of freedom. Specifically, the
low energy effective theory of dense QCD has positive Euclidean
path integral measure, which allows one to establish rigorous
inequalities showing that the color-flavor locked (CFL) phase is
the true vacuum of three flavor, massless QCD. We then describe a
method for simulating dense QCD on the lattice.  We also discuss
applications to electronic systems in condensed matter, such as
generalized Hubbard models.
\end{abstract}

\maketitle

Quark matter is described by a partition function,
\begin{equation}
Z(\mu)=\int {\rm d}A\det \left(M\right)e^{-S(A)},
\label{partition}
\end{equation}
where $\mu$ is the chemical potential for the quark number and
$M=\gamma_E^{\mu}D_E^{\mu}+\mu\gamma_E^4$ is the Euclidean Dirac
operator. Since the Dirac operator $M$ consists of both anti-Hermitian
and Hermitian operators, it has in general complex
eigenvalues. Furthermore, it is not related to its Hermitian
conjugate by any similarity transformations,
\begin{equation}
M=\gamma_E^{\mu}D_E^{\mu}+\mu\gamma_E^4\ne P^{-1}M^{\dagger}P.
\end{equation}
The determinant of $M$ is therefore complex for generic gauge
fields, which has thus far made lattice simulations very difficult
~\cite{Hands:2001jn,Fodor:2001au,Allton:2002zi}.

However, it is shown here that the complexness of the measure of
fermionic dense matter can be ascribed to modes far from the Fermi
surface, which are irrelevant to dynamics at sufficiently high
density in most cases, including quark
matter~\cite{Hong:2002nn,Hong:2003zq}. For modes near the Fermi
surface, there is a discrete symmetry, relating particles and
holes, which pairs the eigenvalues of the Dirac operator to make
its determinant real and nonnegative.

As a simple example, let us consider fermionic matter in 1+1
dimensions, where non-relativistic fermions interact with a gauge
field $A$. The action is in general given as
\begin{eqnarray}
\label{NRA} S\!\!=\!\! \int\!\! d\tau dx~ \psi^{\dagger} \left[ ( -
\partial_\tau + i\phi + \epsilon_F ) - \epsilon( -i\partial_x +
A ) \right] \psi,
\end{eqnarray}
\vspace{-90ex}
\begin{flushright}
     {\parbox[b]{1in}{ \hbox{\tt PNUTP-03/A06}
\hbox{\tt OITS 737} }}
\end{flushright}
\vspace{81ex}
where $\epsilon(p)\simeq p^2/(2m)+\cdots$ is the energy as a function of
momentum.
Low energy modes have momentum near the Fermi points and have energy,
measured from the Fermi points,
\begin{equation}
E(p\pm p_F)\simeq \pm\, v_F p, \quad v_F=\left.{\partial E\over \partial p}\right\vert_{p_F}.
\end{equation}
If the gauge fields have small amplitude and are slowly varying relative to
scale $p_F$, the fast modes are decoupled from low energy physics. The low energy
effective theory involving quasi particles and gauge fields has a positive,
semi-definite determinant.

To construct the low energy effective theory of the fermionic system, we rewrite
the fermion fields as
\begin{equation}
\psi (x, \tau) =\psi_L(x,\tau) e^{+i p_F x} ~+~\psi_R(x,\tau) e^{-i p_F x},
\end{equation}
where $\psi_{L,R}$ describes the small fluctuations of quasiparticles near the
Fermi points.
Using $
e^{\pm ip_F x} ~E( -i\partial_x + A )~ e^{\mp ip_F x} ~\psi (x)
\approx \pm~ v_F (-i\partial_x + A) \psi(x),$
we obtain
\begin{eqnarray}
\label{EFA} S_{\rm eff} &=& \int \, d\tau \, dx~ [
\psi^{\dagger}_L
(-\partial_\tau + i\phi + i\partial_x - A ) \psi_L \nonumber \\
&+&~ \psi^{\dagger}_R ( - \partial_\tau + i\phi - i\partial_x + A ) \psi_R
].
\end{eqnarray}
Introducing the Euclidean (1+1) gamma matrices
$\gamma_{0,1,2}$ and
$\psi_{L,R} = \frac{1}{2} ( 1 \pm \gamma_2) \psi$, we obtain
a positive action:
\begin{equation}
\label{SReff} S_{\rm eff} \!= \!\int d\tau dx\,   \bar{\psi}
\gamma^{\mu} (\partial_{\mu} + iA_{\mu} ) \psi \equiv\!\int d\!\tau
dx\, \bar{\psi} D\!\!\!\!/ \psi .
\end{equation}
Since $D\!\!\!\!/ = \gamma_2{D\!\!\!\!/}^{\dagger}\gamma_2$,
the determinant of $D\!\!\!\!/$ is positive, semi-definite.

In this example, we see that low energy modes near the Fermi
surface can be integrated (in favor of a determinant), leading to
an effective theory without any sign problem whatsoever as long as
they couple to slowly varying background fields. QCD at high
baryon density falls into this category, since the coupling
constant is small at high energy due to asymptotic freedom, and
the corresponding fluctuations in the gauge fields at high density
are small.

A low energy effective theory of QCD at high density, called as
High Density Effective Theory (HDET), has been derived by one of
us~\cite{Hong:1998tn}. (Renormalization group analysis of the
Fermi surface effective theory appears in \cite{Hsu:1998}.)
Consider a quark in one of the patches that cover the Fermi
surface only once. Its momentum can be decomposed as
\begin{equation}
p_{\mu}=\mu v_{\mu}+l_{\mu},\quad |l_{\mu}|<\Lambda\,,~
\Lambda_{\perp}~(\ll\mu),
\end{equation}
where $\Lambda$ and $\Lambda_{\perp}$ are the sizes of patches
perpendicular and parallel to the Fermi surface respectively,
much smaller than the chemical potential but larger than the
scale of interest.
The normalization is enforced by a condition,
\begin{eqnarray}
\sum_{\rm patches}\int_{\Lambda_{\perp}} {\rm d}^2l_{\perp}=4\pi p_F^2.
\end{eqnarray}
The modes near the Fermi surface are given as
\begin{eqnarray}
\psi_+(\vec v_F,x)={1+\vec \alpha\cdot\vec v_F\over2}
e^{-i\mu\vec v_F\cdot \vec x}\psi(x),
\end{eqnarray}
where $\vec \alpha=\gamma_0\vec \gamma$ and $\vec v_F$ is the Fermi velocity
of the modes.
HEDT of quark matter is then described by
\begin{equation}
\label{treeL} {\cal L}_{\rm HDET}=
\bar\psi_+i\gamma_{\shortparallel}^{\mu}D_{\mu}\psi_+-{1\over2\mu}\bar\psi_+
\gamma^0({D\!\!\!\!/}_{\perp})^2\psi_+ ~+~ \cdots,
\end{equation}
where $\gamma^{\mu}_{\shortparallel}=(\gamma^0,\vec v_F\vec v_F\cdot
\vec\gamma)=\gamma^{\mu}-\gamma^{\mu}_{\perp}$.
We see that the leading term has a positive determinant, since
\begin{equation}
M_{\rm eft}=~\gamma^{E}_{\shortparallel}\cdot D(A)~=~
\gamma_5M_{\rm eft}^{\dagger}\gamma_5.
\end{equation}

In order to implement this HDET on lattice, it is convenient to introduce
an operator formalism, where the velocity is realized as an operator,
\begin{equation}
\label{velo} \vec{v} =   \frac{-i }{\sqrt{- \nabla^2}}
~\frac{\partial}{\partial \vec{x}}~~.
\end{equation}
Then the quasi quarks near the Fermi surface become
\begin{equation}
\psi = \exp \left( + i \mu x \cdot v \right)
{1+ \alpha \cdot v\over 2}
\psi_+  .
\end{equation}
Now, neglecting the higher order terms,
the Lagrangian becomes with $X=\exp(i\mu x\cdot v)
(1+\alpha\cdot v)/2$,
\begin{eqnarray}
\label{leading2}
{\cal L}_{\rm HDET} =  \bar{\psi}_+  \gamma^\mu_\parallel
\left(
\partial^\mu + i A^\mu_+ \right) \psi_+,
\end{eqnarray}
where $A^\mu_+  =  X^{\dagger}\,A^\mu\,X$ denotes soft gluons whose
momentum $|p_{\mu}|<\mu$.
Since $v \cdot \partial \, v \cdot \gamma =
\partial \cdot \gamma~$, we get
$\gamma^\mu_{\parallel} \partial^\mu = \gamma^\mu \partial^\mu$,
which shows that the operator formalism automatically covers modes
near the full Fermi surface.

Integrating out the fast modes, modes far from the Fermi surface and
hard gluons, the QCD partition function~(\ref{partition}) becomes
\begin{equation}
Z(\mu)=\int {\rm d}A_+\det \left(M_{\rm eff}\right)e^{-S_{\rm eff}(A_+)},
\label{epartition}
\end{equation}
where
\begin{eqnarray}
S_{\rm eff}=\int_{x_E}\left({1\over4}F_{\mu\nu}^aF_{\mu\nu}^a +{M^2\over
16\pi}A_{\perp\mu}^{a}A_{\perp\mu}^{a}\right)
+\cdots
\end{eqnarray}
and $A_{\perp}=A-A_{\parallel}$, the Debye mass
$M=\sqrt{N_f/(2\pi^2)}g_s\mu$\,. At high density the higher order
terms ($\sim \Lambda/\mu$) are negligible and the effective action
becomes positive, semi-definite. Therefore, though it has
non-local operators, HDET in the operator formalism is free from
the sign problem and can be used to simulate the Fermi surface
physics like superconductivity. Furthermore, being exactly
positive at asymptotic density, HDET allows to establish rigorous
inequalities relating bound state masses and forbidding the
breaking of vector symmetries, except baryon number, in dense
QCD~\cite{Hong:2003zq}.

With the help of the previous two examples, we propose a new way
of simulating dense QCD, which evades the sign problem.
Integrating out quarks far from the Fermi surface, which are
suppressed by $1/\mu$ at high density, we can expand the
determinant of Dirac operator at finite density,
\begin{equation}
\det \left(M\right) = \left[ {\rm real, positive} \right] \left[
1 ~+~ {\cal O} \left( \frac{ {{\bf F}}  }{\mu^2} \right) \right].
\end{equation}
As long as the gauge fields are slowly varying, compared to
the chemical potential $\mu$, the sign problem can be evaded.
As a solution to the sign problem,
we propose to use two lattices with different spacings, a finer lattice with
a lattice spacing  ${a_{\rm det}}\sim \mu^{-1}$ for fermions and a coarser lattice
with a lattice spacing  ${a_{\rm gauge}}\ll \mu^{-1}$ for
gauge fields and then compute the determinant on such  lattices.

\begin{figure}[htb]
\includegraphics[width=17pc]{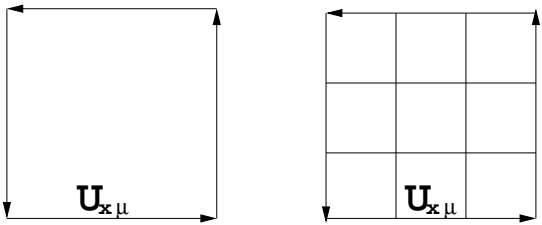}
\caption{Simulation with two lattices with different lattice spacings }
\label{fig}
\end{figure}

The determinant is a function of plaquettes ${\bf \{ U_{x \mu}
\}}$ which are obtained by interpolation from the plaquettes on
the coarser  lattice with spacing $a_{\rm gauge}$. To get the link
variables for the finer lattice, we interpolate the link variables
${\bf U_{x \mu}} \in SU(3)$ (see Fig.~\ref{fig}): Connect any two
points $g_1, g_2$ on the group manifold as
\begin{equation}
g(t) = g_1 + t (g_2 - g_1)~,~ 0 \leq t \leq 1\,.
\end{equation}


For importance sampling in the lattice simulation,
one can use  the leading part of the
determinant, $[{\rm real}, {\rm positive}]$.
This proposal provides a nontrivial check on analytic results at
asymptotic density and can be used to extrapolate to intermediate
density. Furthermore, it can be applied  to condensed matter systems
like High-$T_c$ superconductors, which in general suffers from a
sign problem.

\bigskip
{\bf Acknowledgement}

The work of D.K.H. is supported by KOSEF grant number
R01-1999-000-00017-0 and also by Pusan National University
Research Grant, $1999\sim2003$. The work of S.H. was supported in
part under DOE contract DE-FG06-85ER40224 and by the NSF through
through the USA-Korea Cooperative Science Program, 9982164.

\bigskip

\end{document}